\documentclass[twocolumn, pra, amssymb, superscriptaddress, aps, showpacs,preprintnumbers,amsmath,floatfix]{revtex4} 
\usepackage{multirow}
\usepackage{isotope}
\usepackage{graphicx}
\usepackage{color}

\newcommand{\ket}[1]{\left| #1 \right\rangle}

\begin{document}

\title{Internal conversion from excited electronic states of $\isotope[229]{Th}$ ions }

\author{Pavlo V. \surname{Bilous}}
\email{Pavlo.Bilous@mpi-hd.mpg.de}
\affiliation{Max-Planck-Institut f\"ur Kernphysik, Saupfercheckweg 1, D-69117 Heidelberg, Germany}

\author{Georgy A. \surname{Kazakov}}
\affiliation{Vienna Center for Quantum Science and Technology (VCQ), Atominstitut, TU Wien, Stadionallee 2, 1020 Vienna, Austria}

\author{Iain D. \surname{Moore}}
\affiliation{Department of Physics, University of Jyv\"askyl\"a, Survontie 9, FI-40014 Jyv\"askyl\"a, Finland}

\author{Thorsten \surname{Schumm}}
\affiliation{Vienna Center for Quantum Science and Technology (VCQ), Atominstitut, TU Wien, Stadionallee 2, 1020 Vienna, Austria}

\author{Adriana P\'alffy}
\email{Palffy@mpi-hd.mpg.de}
\affiliation{Max-Planck-Institut f\"ur Kernphysik, Saupfercheckweg 1, D-69117 Heidelberg, Germany}


\date{\today}

\begin{abstract}
The process of internal conversion from excited electronic states is investigated theoretically for the case of the vacuum-ultraviolet nuclear transition of $\isotope[229]{Th}$. Due to the very low transition energy, the $\isotope[229]{Th}$ nucleus offers the unique possibility 
to open  the otherwise forbidden internal conversion nuclear decay channel for thorium ions via optical laser excitation of the electronic shell. We show that this  feature can be exploited to  investigate the isomeric state properties via observation of internal conversion from excited electronic configurations of $\isotope{Th}^+$ and $\isotope{Th}^{2+}$ ions. A possible experimental realization of the proposed scenario at the nuclear laser spectroscopy  facility IGISOL in Jyv\"askyl\"a, Finland is discussed.

\end{abstract}


\maketitle


\section{Introduction}
The $\isotope[229]{Th}$ isotope is unique throughout the entire nuclear chart due to its first nuclear excited state lying energetically in the optical  range~\cite{Beck}. In the language of nuclear physics, this state is an isomer, i.e., a long-lived excited nuclear state. At present the most accepted value for the energy of this level is  $E_{\mathrm{m}}=7.8 \pm 0.5$~eV~\cite{Beck}, rendering it accessible to vacuum ultra-violet (VUV) lasers. Due to the very low transition energy, the radiative decay of the state is strongly suppressed leading to an exceptionally long lifetime of the isomer. The ratio of  radiative width to transition energy  is presently estimated at $\gamma_{\gamma}/E_{\mathrm{m}}=10^{-19}$. Given this very narrow width and the high robustness of nuclei to external perturbations \cite{clock_campbell_2012}, the isomeric state has been proposed for novel applications such as a nuclear frequency standard with unprecedented accuracy \cite{clock_peik_2003,clock_campbell_2012,clock_peik_2015} or a nuclear laser~\cite{nucl_laser}. On the other hand, these features render a direct spectroscopic search for the nuclear resonance with a narrow-band laser extremely tedious. A coarse determination of the isomer energy or a restriction of the search range would therefore be extremely helpful.

Since the nuclear transition energy lies above the first ionization threshold of Th (see Table~\ref{Ionpot}), the decay of the nuclear state can occur not only radiatively but also via internal conversion (IC). In the process of IC, the nuclear transition energy is transferred to a bound electron in the atom, which is then ionized. In the case of $\isotope[229]{Th}$, the 7.8~eV can be transferred to one of the  valence electrons, which then leaves the atom. This decay channel is much stronger than the radiative decay, with an IC coefficient, i.e., the ratio of the IC rate to the radiative decay rate of the same transition, of  $\alpha=\gamma_{IC}/\gamma_{\gamma}\sim 10^9$ \cite{karpeshin2007,tkalya_prc2015}. 

Typically, in the process of IC the initial electronic state is the ground state of the atom.  Due to the very low-lying nuclear excited state, $\isotope[229]{Th}$ is a unique candidate for investigating IC of electrons from an excited electronic state. Optical or VUV lasers can be used to excite the atomic shell and maintain the outermost electrons available for IC in excited states. This opportunity exists and is relevant only for $\isotope[229]{Th}$, since for higher nuclear transition energies mostly electrons from inner shells of the ground-state electronic configuration are ionized. Practically, inner-shell vacancies are notoriously short-lived and excited electronic configurations cannot be maintained long enough for IC to occur. 
In this work we investigate this unique feature of  $\isotope[229]{Th}$ to design  new possible means to detect the isomeric state via observation of IC from excited electronic configurations.

Better knowledge of the isomeric state energy and its properties is required for all nuclear quantum optics applications based on the $\isotope[229]{Th}$ isomeric transition  proposed so far. Several approaches have been pursued, for example atomic spectroscopy of trapped Th ions~\cite{campbell_prl2011} and laser spectroscopy of Th nuclei grown in VUV-transparent crystal lattices~\cite{Rel10,stellmer_doped, stellmer_radiolum}.  Significant progress has been recently achieved by the detection of the IC electrons from the isomeric state in neutral $\isotope[229]{Th}$ atoms~\cite{lars_nature}. This result is considered to be the first direct observation of the low-energy isomeric state of $\isotope[229]{Th}$. However, direct or indirect excitation of the isomeric state has not been achieved so far~\cite{synchro83,yamaguchi}, raising the question of whether the isomeric energy lies somewhat higher than the region thus far considered~\cite{synchro83,tkalya_prc2015}. An experimental confirmation of the $\isotope[229\mathrm{m}]{Th}$ isomeric state energy  and lifetime thus remains a very important experimental challenge.

\begin{figure}[ht!]
\centering
\includegraphics[width=0.48\textwidth]{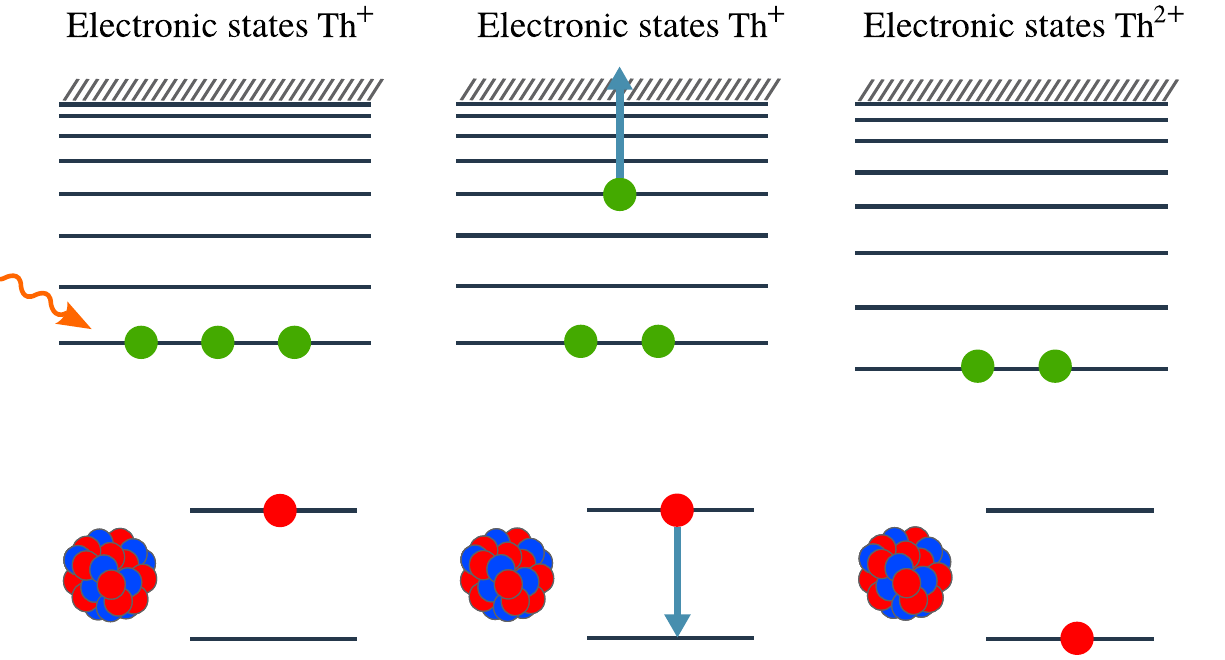}
\caption{(Color online.) A schematic illustration of IC from an excited electronic state of  $\isotope{Th}^+$ leading to the formation of $\isotope{Th}^{2+}$ ions. The generic electronic level scheme of the outermost three electrons is depicted in the upper part of the figure, while the lower part illustrates the ground and isomeric nuclear states. Photoexcitation creates the excited electronic state in  $\isotope{Th}^+$ (left graph). This opens the IC channel for the isomeric state, such that an electron is ionized (middle graph) producing $\isotope{Th}^{2+}$ ions (right graph). \label{fig1} }
\end{figure}

Here we envisage a new possible approach to detect the isomeric state taking advantage of IC from excited electronic states.  IC is energetically allowed in the neutral atom, but should be forbidden in  Th ions, as shown by the ionization potential values tabulated in Table~\ref{Ionpot}. Thus, IC  from the electronic ground state becomes energetically forbidden for higher degrees of ionization  so that {\it only} IC from excited states remains allowed. 
If a $\isotope[229\mathrm{m}]{Th}$ ion prepared in an excited electronic state undergoes IC at a rate fast enough compared with the spontaneous decay of the electronic excited state,  ions of a higher charge state will be produced in the process. Detection of such ions  can be performed with close-to-unity efficiency and direct comparison with the case of $\isotope[232]{Th}$ (which does not possess any nuclear states at optical energies) will indicate the presence of the isomeric state. Based on the observation of occurring IC-induced ionization, one can  estimate the isomeric energy and  compare the IC rate with theoretical predictions. A schematic illustration of the IC from an excited electronic state of  $\isotope{Th}^+$ leading to the formation of $\isotope{Th}^{2+}$ ions is presented in Fig.~\ref{fig1}.

\begin{table}[h!]
\centering
\begin{tabular}{|c|c|c|c|c|c|}
\hline
Ion charge & 0 & 1+ & 2+ & 3+ & 4+\\
\hline
Ion. threshold (eV) & 6.3 & 12.1 & 20.0 & 28.7 & 58\\
\hline
\end{tabular}
\caption{\label{Ionpot}Ionization thresholds for several $\isotope{Th}$ ions~\cite{nist,michigan}.}
\end{table}

Following present speculations \cite{synchro83,tkalya_prc2015},  in this work we analyze possible excited states as candidates for the initial IC configuration assuming that the value of $E_\mathrm{m}$ may also lie above $7.8$~eV. As upper limit we choose  $20.0$~eV  which is the ionization potential of $\isotope{Th}^{2+}$. Comparing the IC rates and the rates of radiative decay for the electronic levels, we show that this detection scheme is applicable for  $\isotope{Th}^+$  in the case when $E_{\mathrm{m}}$ lies between approx.~$9.0$ and $12.0$~eV, while $\isotope{Th}^{2+}$ can be used only in the rather unlikely case that $E_\mathrm{m}$ is higher than $19.1$~eV, but less than $20.0$~eV. A possible experimental verification of this scenario at the IGISOL facility in Jyv\"askyl\"a, Finland is discussed towards the end of the paper.

The paper is structured as follows. We start by giving a short overview of the theoretical calculations for IC rates from ground and excited state electronic configurations in Section~\ref{theory}. Numerical results are presented in Section \ref{numres}. In the following Section we discuss possible experimental approaches to reach excited electronic states relevant for the IC scenario, with the proposed experiment for isomeric state detection in Section \ref{measurement}. The paper closes with a summary in Section \ref{summary}. Atomic units ($m_e=\hbar=e=1$) are used throughout the paper unless otherwise mentioned.

\section{Theoretical Considerations \label{theory}}
IC is one of the few nuclear processes directly involving atomic electrons. The electronic and nuclear degrees of freedom couple electromagnetically, and the interaction Hamiltonian for nuclear transitions of electric multipolarities
is determined by  the Coulomb interaction between the nucleus and the converted electron,
\begin{equation}
H_{\mathrm{el}} = \int d^3r_n\frac{\rho_n(\vec{r}_n)}{|\vec{r}_e-\vec{r}_n|}\, ,
\label{hel}
\end{equation} 
where $\rho_n(\vec{r}_n)$ is the nuclear charge density,  $\vec{r}_n$ ($\vec{r}_e$) denotes the nuclear (electronic) coordinate and
the integration is performed over the whole nuclear volume. For magnetic nuclear transitions  the  Hamiltonian is given in second order perturbation theory by the interaction between the electronic and nuclear charge currents \cite{Adriana.PRA73},
\begin{equation}
H_{\mathrm{magn}} = - \frac{1}{c} \vec{\alpha} \int d^3r_n \frac{\vec{j}_n(\vec{r}_n)}{|\vec{r}_e-\vec{r}_n|},
\label{hmagn}
\end{equation}
where $c$ is the speed of light,   $\vec\alpha$ is the vector of the Dirac
$\alpha$ matrices $(\alpha_x,\alpha_y,\alpha_z)$  and $\vec{j}_n(\vec{r}_n)$ is the nuclear current.
Also in this case the integration is performed over the whole nuclear volume. The isomeric transition in $\isotope[229]{Th}$ is a multipole mixture between a stronger $M1$ (magnetic dipole) transition and a typically neglected $E2$ (electric quadrupole) transition. Indeed, for the pure radiative decay the contribution of the electric quadrupole part to the transition rate is completely negligible due to the very small nuclear transition energy. However, this might not hold true for IC rates which have a weaker dependence on the transition energy.

The IC rate for a specific multipolarity is determined by the squared absolute value of the interaction Hamiltonian matrix element, usually averaged over all possible initial states and summed over all final states. The total IC rate for a transition with multipole mixing is then given by the sum of the individual $M1$ and $E2$ IC rates. 
For the simplest case of a single bound electron undergoing IC in an otherwise closed-shell electronic configuration, the IC rate for the $M1$ contribution yields 
\begin{eqnarray}\label{gammaicM1}
\Gamma^{M1}_{\mathrm{IC}} &=& 
\frac{8 \pi^2}{9}B_\downarrow(M1) \sum_{\kappa}(2j+1)\nonumber \\
&\times&
 (\kappa_i+\kappa)^2
\begin{pmatrix}
j_i & j & 1 \\
1/2 & -1/2 & 0
\end{pmatrix}^2
|R_{\varepsilon\kappa}^{\mathrm{M1}}|^2\, ,
\end{eqnarray}
and correspondingly for the $E2$ contribution 
\begin{eqnarray}
\label{gammaicE2}
\Gamma^{E2}_{\mathrm{IC}} &=& 
\frac{8 \pi^2}{25}B_\downarrow(E2) \sum_{\kappa}(2j+1)\nonumber \\
&\times&
\begin{pmatrix}
j_i & j & 1 \\
1/2 & -1/2 & 0
\end{pmatrix}^2
|R_{\varepsilon\kappa}^{\mathrm{E2}}|^2\, .
\end{eqnarray}
In the equations above,  $j$ ($j_i$) and $\kappa$ ($\kappa_i$) are the total angular and the Dirac angular momentum quantum numbers for the continuum (initial bound) electron and 
\begin{equation}
B_\downarrow(M1/E2) = \frac{
|\left\langle I_g \right\|
\hat{\mathcal{M}}/\hat{\mathcal{Q}}
\left\| I_e \right\rangle |^2
}
{2I_e+1}
\end{equation}
denotes the reduced probability of the nuclear transition from the isomeric to the ground state. The notations $I_e$ and $I_g$ represent the isomeric and ground state nuclear spin, respectively, and $\hat{\mathcal{M}}$ ($\hat{\mathcal{Q}}$) is the nuclear magnetic dipole (electric quadrupole) operator \cite{RingSchuck}. The summation over the   continuum partial wave Dirac quantum number $\kappa$ is performed such that the selection rules for the particular multipolarity transition apply.  The radial integrals $R_{\varepsilon\kappa}^{\mathrm{M1}}$ and $R_{\varepsilon\kappa}^{\mathrm{E2}}$ in Eqs.~(\ref{gammaicM1}) and (\ref{gammaicE2}) are given by 
\begin{eqnarray}
R_{\varepsilon\kappa}^{\mathrm{M1}} &=& \int_{0}^{\infty} dr \Bigl( g_{n_i \kappa_i}(r)f_{\varepsilon \kappa}(r) + g_{\varepsilon \kappa}(r) f_{n_i\kappa_i}(r) \Bigr)\, , \\
R_{\varepsilon\kappa}^{\mathrm{E2}} &=& \int_{0}^{\infty} \frac{dr}{r} \Bigl( g_{n_i \kappa_i}(r)g_{\varepsilon \kappa}(r) + f_{\varepsilon \kappa}(r) f_{n_i\kappa_i}(r) \Bigr)\, ,
\end{eqnarray}
where $g_{\beta \kappa}$ and $f_{\beta \kappa}$ are the radial wave functions of the initial (bound) and final (continuum) one-electron states, respectively. The total wave function for the electron is given by
\begin{equation}
\ket{\beta \kappa m} =
\begin{pmatrix}
g_{\beta \kappa}(r) \Omega_{\kappa m}(\hat{r}) \\
if_{\beta \kappa}(r) \Omega_{-\kappa m}(\hat{r})
\end{pmatrix}\;,
\end{equation}
where the functions $\Omega_{\kappa m}(\hat{r})$ of the argument $\hat{r}=\vec{r}/r$ are the spherical spinors. The generic notation $\beta$ stands for the principal quantum number $n$ for bound electron orbitals and for the continuum electron energy $\varepsilon$ for free electron wave-functions, respectively. 

The IC rate expressions in the case of Th ions are more complicated generally speaking due to possible electronic couplings between the electrons in the outer-shell orbitals with principal quantum numbers $n=5,6,7$. These shells are not closed and one should in this case consider (i) the coupling of the angular momenta of the IC electron and the remaining two valence electrons of the initial ion, (ii) the coupling of the angular momenta of the free electron and the electronic shell of the final ion, (iii) the possibility of simultaneous population of different electronic states of the final ions.

An accurate treatment of the case of two electrons in the outer electronic shell (as in the case of $\isotope{Th}^{2+}$)  leads again to the expressions (\ref{gammaicM1}) (for  $M1$ transition multipolarity) and  (\ref{gammaicE2}) (for  $E2$ transition multipolarity) and the corresponding selection rules for the IC electron parity. However, now the IC rate is non-zero only if the total angular momentum of the spectator electron remains unchanged before and after conversion. A further increase of the number of outer-shell electrons makes the angular momenta coupling more complex as different coupling schemes are rendered possible. For three electrons in the outer electron shell (the case of $\isotope{Th}^{+}$), we distinguish between two cases. First, we consider a configuration $\ket{ J_iM_i}$ of electrons with individual total angular momenta $j_1$, $j_2$, $j_i$ (the latter denoting the electron which undergoes IC) such that
\begin{itemize} 
\item $j_1$, $j_2$ are first coupled to the momentum $J_0$.
\item $J_0$ and $j_i$ are then coupled to the total momentum $J_i$.
\end{itemize}
Under these conditions the initial coupling to the momentum $J_0$ is not broken by  IC and similar to the case of two electrons, we derive the expressions (\ref{gammaicM1}) and  (\ref{gammaicE2})  with the corresponding restrictions on the IC electron parity. Further selection rules require the angular momenta $j_1$, $j_2$ of the non-IC electrons and their common momentum $J_0$ to be conserved. The total angular momentum $J_f$ of the final two-electron configuration should therefore exactly equal $J_0$.

Now let us consider a configuration $\ket{ J_iM_i}$ of electrons with total angular momenta $j_1$, $j_2$, $j_i$ such that
\begin{itemize} 
\item $j_2$, $j_i$ are first coupled to the momentum $J_0$.
\item $J_0$ and $j_1$ are then coupled to the total momentum $J_i$.
\end{itemize}
In this case the initial coupling to the momentum $J_0$ is broken by IC. It leads to the expressions (\ref{gammaicM1}) (for  $M1$ transition multipolarity) and  (\ref{gammaicE2}) (for  $E2$ transition multipolarity)  with the corresponding restrictions on the IC electron parity, but with an additional  coefficient
\begin{equation}\label{Lambda}
\Lambda = (2J_0+1)(2J_f+1)
\left\{
\begin{matrix}
j_1 & j_2 & J_f \\
j_i & J_i & J_0
\end{matrix}
\right\}^2\, .
\end{equation}
The selection rules still imply conservation on the angular momenta $j_1$ and $j_2$. The other restrictions are contained in the $6j$-symbol, namely, the four sets $(j_1,j_2, J_f)$, $(j_1,J_i, J_0)$, $(j_1,j_2 ,J_0)$, $(j_i,J_i, J_f)$ have to satisfy the triangle rule. Note that in this case the momentum $J_f$ is not restricted only to the value $J_0$. 

\section{Numerical results \label{numres}}
We have calculated the total IC rate for a number of excited electronic configurations of Th in the charge states $\isotope{Th}^{+}$ and $\isotope{Th}^{2+}$. The key ingredients in the calculation of the IC rate (\ref{gammaicM1}), (\ref{gammaicE2}) and (\ref{Lambda}) are
\begin{itemize}
\item the reduced probabilities of the nuclear transition $B_\downarrow(M1)$ and $B_\downarrow(E2)$. At the moment, only theoretical estimates are available for the nuclear transition strength and the value of $B_\downarrow$ can vary over a wide range depending on the method used for its calculation~\cite{tkalya_prc2015}. Here we assume $B_\downarrow (M1) = 0.048\; \mathrm{ W.u.}$~ (Weisskopf units) in accordance to Ref.~\cite{tkalya_jetp}. For the electric quadrupole transition we consider a recent preliminary estimate $B_\downarrow (M1) = 29\; \mathrm{ W.u.}$ \cite{Minkov2016}.
\item the radial wave functions $g_{n_i \kappa_i}$ and $f_{n_i \kappa_i}$ for the bound electron. 
\item the radial wave functions $g_{\varepsilon \kappa}$ and $f_{\varepsilon \kappa}$ for the continuum electron for all $\kappa$ values allowed by the selection rules.
\item the electronic energy levels for the thorium ions before and after IC.
\end{itemize}

We evaluate the bound electron radial wave functions $g_{n_i \kappa_i}$ and $f_{n_i \kappa_i}$  with the grasp2K package provided by the
Computational Atomic Structure Group~\cite{grasp}. The package consists of a number of tools for computing
relativistic wave functions, energy levels, transition rates and other properties of many-electron atoms. 
 The energy levels for many-electron heavy ions such as $\isotope{Th}$ (atomic number $Z=90$) calculated with grasp2K are not reliable on the degree of accuracy required here.  Although the relative accuracy is better than $10^{-5}$, due to the very large total electronic energy, errors on the order of eV may occur and  often the order of the levels is uncertain.  This can considerably lower the quality of the calculated atomic level scheme.  We therefore choose to adopt  the initial- and final-state electronic energies from the experimental database~\cite{dblevels}. The experimental values were obtained in laser spectroscopy laboratories and are accurate on the level of   10$^{-3}$~cm$^{-1}$, which is  beyond reach for atomic structure calculations. In contrast, the electronic wave functions can be calculated with grasp2K with sufficient accuracy. 

For evaluation of the radial wave functions $g_{\varepsilon \kappa}$ and $f_{\varepsilon \kappa}$ of the con\-ti\-nu\-um state of the IC electron we use the program \textit{xphoto} from the Ratip package~\cite{ratip}. \textit{Xphoto}  performs relativistic calculations for photo-ionization  cross-sections of multi-electron atomic systems and can provide electronic wave functions of the continuum states for all required $\kappa$ values.

\begin{figure*}[ht!]
\centering
\includegraphics[width=16cm]{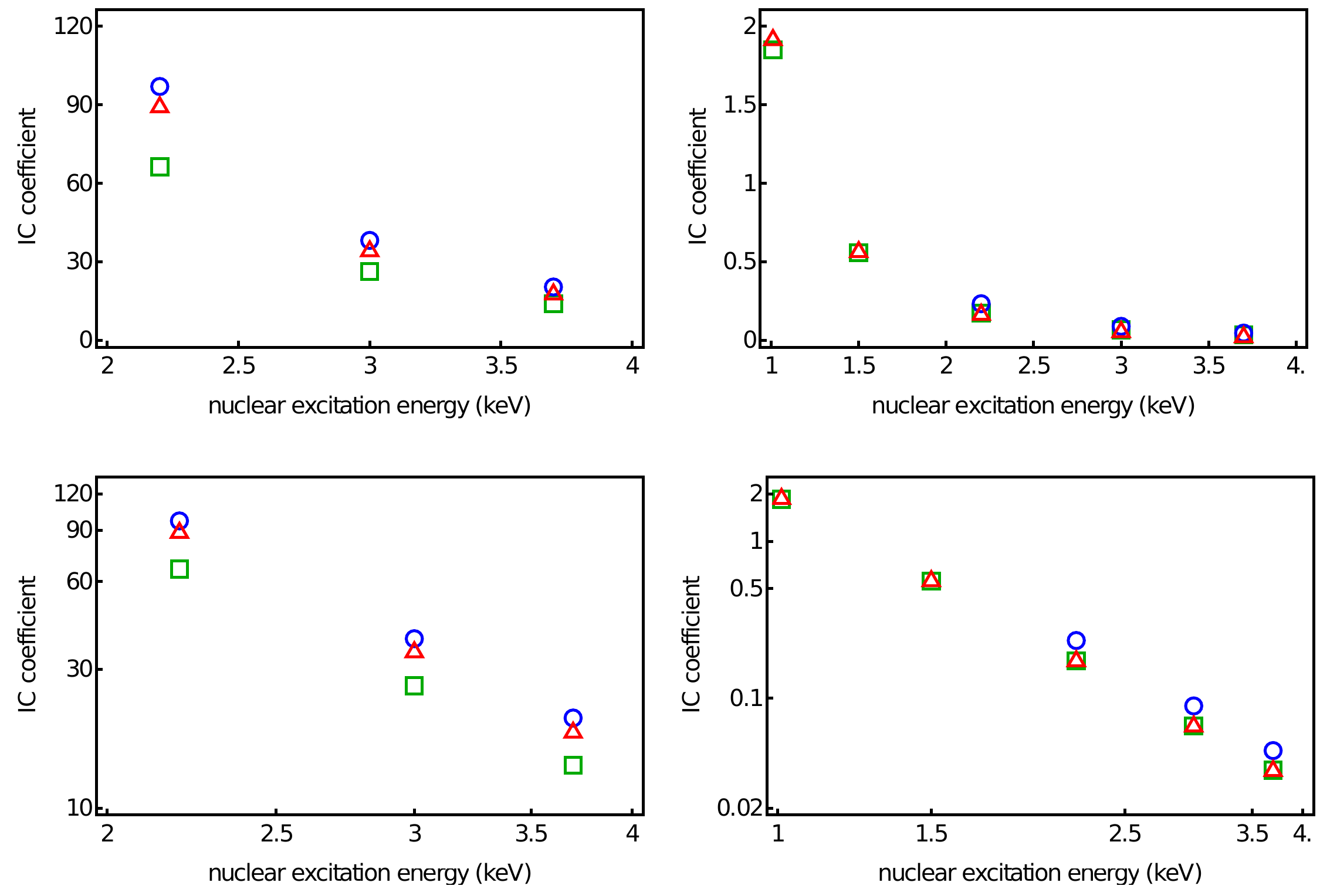}
\caption{(Color online.) IC coefficients for the $7s$-electron (the left two graphs) and $6d$-electron (the right two graphs) of thorium calculated in the present work (red triangles), in Ref.~\cite{trzhaskowskaya} (green squares) and in Ref.~\cite{pauli} (blue circles) as a function of nuclear transition energy. The lower graphs present the same values on a logarithmic scale to check the expected linear slope in the IC coefficient \cite{pauli}. \label{fig2}}
\end{figure*}

We have validated our calculation method and numerical results by comparing IC coefficients with theoretical values tabulated in the literature \cite{pauli,trzhaskowskaya} for a number of test cases. The compilations in Refs.~\cite{pauli,trzhaskowskaya} 
use the relativistic Dirac-Hartree-Fock method for the calculation of electronic wavefunctions. In addition, Ref.~\cite{trzhaskowskaya}  takes into account the effect of atomic vacancies created in the conversion process. 
In  Fig.~\ref{fig2} we depict the calculated IC coefficients as a function of the nuclear transition energy together with data from Refs.~\cite{pauli, trzhaskowskaya}. We consider a fictitious nuclear transition of multipolarity $M1$ which ionizes the $6d$- and $7s$-electrons in the neutral Th atom. In all cases good agreement with at least one of the tabulated values is achieved.

\begin{figure}[ht!]
\centering
\includegraphics[width=9cm]{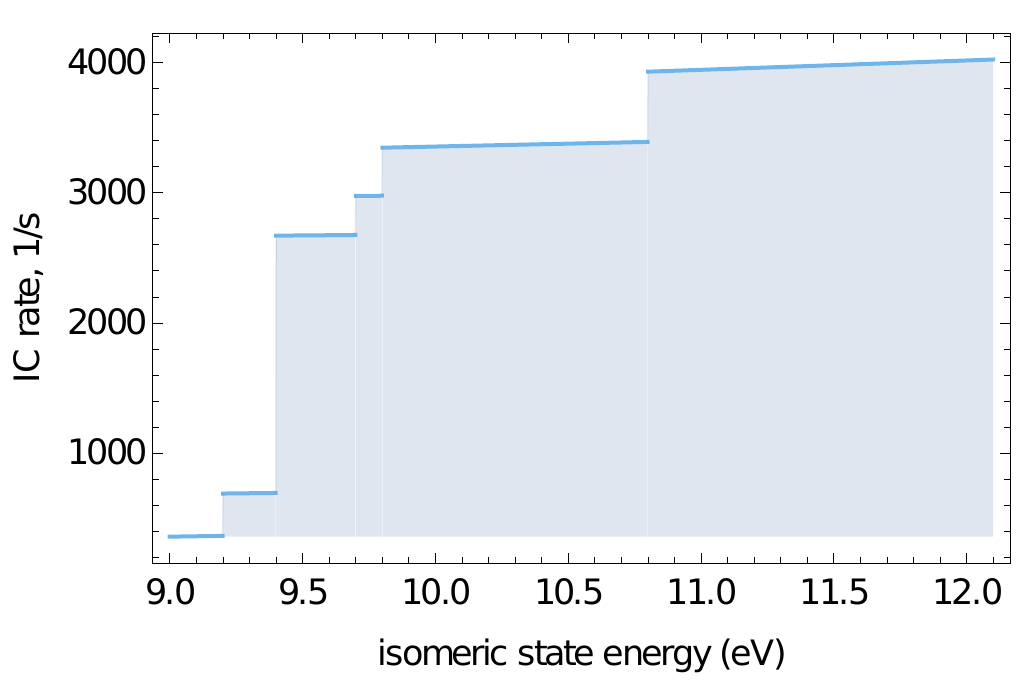}
\caption{ Internal conversion rates for the state $5f6d_2$ at 30223~$\mathrm{cm}^{\mathrm{-1}}$ of $~^{229\mathrm{m}}\mathrm{Th}^+$ as a function of the possible isomeric transition energy. For comparison, the half-life of the electronic excited state was calculated to be 0.4 s.\label{fig3}}
\end{figure}

We now proceed with numerical results for IC from  excited electronic states of $\isotope{Th}^{+}$ and $\isotope{Th}^{2+}$. We take into account both the $M1$ contribution and the multipole mixing of the $E2$ multipolarity contribution, which although negligible for IC of the thorium ground state \cite{tkalya_prc2015}, turns out to be relevant for IC from excited states.  In the following tables we list the calculated total IC rates for a number of relevant cases and consider  a larger interval for the isomeric energy $E_{\mathrm{m}}$. The minimum isomeric energy required for the respective IC channel to be open is labeled as $E_\mathrm{m}^{\mathrm{min}}$. In particular we focus on $\isotope{\text{Th}}^\text{+}$ configurations and two $\isotope{\text{Th}}^\text{2+}$ excited states that are of  interest for the experimental investigation of the electron bridge process~\cite{bridge_exc}. We present the IC rate as a function of the possible nuclear isomeric state energy and take into account all available final states within the considered energy range. Note that the number of digits to represent the IC rates is chosen in each case to be sufficient to depict the dependence of the result on the isomeric state energy within the framework of the calculation model. With increasing charge state, the required electronic excitation and ionization energies are also increasing. Thus, we note that due to the large ionization potential of  $\isotope{Th}^{2+}$, the required isomeric energy to render IC from the chosen excited electronic configuration is approx.~19 eV, which is according to present knowledge highly unlikely. We conclude that the most interesting ionization state for IC from an excited electronic state is $\isotope{\text{Th}}^\text{+}$.

\subsection{Results for $\isotope{Th}^{+}$}
Table~\ref{lifetimes} presents  theoretical radiative lifetimes calculated with the grasp2K package for the excited electronic states of interest for the IC scheme. We note that the Auger decay of these states is energetically forbidden. 
 The choice of the excited states is unfortunately limited by the available electronic energy data in the experimental database \cite{dblevels} and by the short radiative lifetimes of the excited states. 
Based on the values in Table~\ref{lifetimes}, we choose to consider two suitable initial excited electronic states: $5f6d_2$ at 30223 $\text{cm}^{\text{-1}}$ with $J=15/2$ and the $7s_27p$ at 31626 $\text{cm}^{\text{-1}}$ with $J=1/2$. 
The IC rates  for these two configurations  are presented in Tables~\ref{5f6d2} and \ref{7s27p}. The excited state $5f6d_2$  has no electric-dipole decay channels; its calculated radiative lifetime is 0.4 s.  IC  from the $5f6d_2$ state becomes possible provided that the isomeric state energy would be higher than $9.0\text{ eV}$, in which case the characteristic decay time  becomes considerably less than the level lifetime.  Provided that the isomeric energy indeed lies higher than 7.8 eV as speculated at present, this opportunity seems to be unique, as the other states at such high excitation energies typically decay very fast (see Table~\ref{lifetimes}, the case for the $7s_27p$ state at 31626 $\text{cm}^{\text{-1}}$). On the other hand, in case the isomer energy is higher than 12 eV, the excitation method would no longer be applicable since the ground state $6d_27s$ electrons could undergo IC. We are therefore focusing on the range of $9.0$~eV to $12.0$~eV isomeric state energy. 
We note here that the laser excitation of this particular electronic configuration is not straightforward due to the large difference of its total angular momentum and the angular momentum of the ground state: $\frac{15}{2}-\frac{3}{2}=6$. Several possible experimental approaches are addressed in the next Section.

A graphical representation of the values in Table~\ref{5f6d2} is given in  Fig.~\ref{fig3}. 
We depict the IC rates for the initial $5f6d_2$ state  as a function of the possible nuclear isomeric state energy and take into account all available final states within the considered energy range.  The step-like jumps represent openings of new channels of decay corresponding to different electronic states of the final ion $\isotope{Th}^{2+}$.

\begin{table}[H]
\centering
\begin{tabular}{|c|c|c|c|}
\hline
Ion charge & Configuration & Energy, $\text{cm}^{-1}$ & Lifetime \\
\hline
\multirow{2}{*}{1+} & $5f6d_2$ & 30223 & 0.4 s\\
\cline{2-4}
&$7s_27p$ & 31626 &  40 ns\\
\hline
\multirow{2}{*}{2+} & $5f7s$ & 7501 &   20 $\mu$s\\[0.1cm]
\cline{2-4}
&$6d7s$ & 16038 & 100  ns\\
\hline
\end{tabular}
\caption{\label{lifetimes} Lifetimes of the considered excited electronic states.}
\end{table}

\begin{table}[ht!]
\centering
\begin{tabular}{|c|c||c||c|c|c|}
\hline
\multicolumn{2}{|c||}{Final state} &
\multirow{2}{*}{$E_\mathrm{m}^{\mathrm{min}}$,~eV} &
\multirow{2}{*}{$E_\mathrm{m}$,~eV} &
\multicolumn{2}{c|}{Rate, $\mathrm{s}^{-1}$} \\
\cline{1-2}
\cline{5-6}
Config. & $E$, $\mathrm{cm}^{-1}$ & &  & Pure M1 &  M1+E2 \\
\hline
\multirow{11}{*}{$6d_2$} &
      \multirow{6}{*}{6538}   
             &
              \multirow{6}{*}{9.2}
               &  9.2 & 284 & 325 \\
&&         &  9.5 & 282 & 323 \\
&&         &  10.0 & 279 & 319 \\
&&         &  10.5 & 276 & 315 \\
&&         &  11.0 & 273 & 311 \\
&&         &  11.5 & 270 & 308 \\
      \cline{2-6}
&   \multirow{5}{*}{10543}   
             &
              \multirow{5}{*}{9.7}
               &  9.7  & 258 & 301 \\
&&         &  10.0 & 256 & 298 \\
&&         &  10.5 & 252 & 293 \\
&&         &  11.0 & 249 & 289 \\
&&         &  11.5 & 246 & 286 \\
\cline{1-6}
 \multirow{19}{*}{$5f6d$}
      & \multirow{6}{*}{4490}
              & \multirow{6}{*}{9.0}
               & 9.0 & 253 & 356 \\
 &&        & 9.5 & 264 & 373 \\
 &&        & 10.0 & 275 & 389 \\
 &&        & 10.5 & 284 & 404 \\
  &&        & 11.0 & 294 & 419 \\
   &&        & 11.5 & 302 & 432 \\
      \cline{2-6}
&           \multirow{5}{*}{8437}   
             &
              \multirow{5}{*}{9.4}
               &  9.4  & 1541 & 1976 \\
&&         &  10.0 & 1535 &     1972 \\
&&         &  10.5 & 1531 &     1969 \\
&&         &  11.0 & 1527 &     1966 \\
&&         &  11.5 & 1524 &     1964 \\
      \cline{2-6}
&        \multirow{5}{*}{11277}   
             &
              \multirow{5}{*}{9.8}
               &  9.8  & 238 & 368\\
&&         &  10.0 & 245 &     376 \\
&&         &  10.5 & 260 &     395 \\
&&         &  11.0 & 273 &     413 \\
&&         &  11.5 & 286 &     429 \\
      \cline{2-6}
&         \multirow{3}{*}{19010}   
             &
              \multirow{3}{*}{10.8}
               &  10.8 & 368 & 539 \\
&&         &  11.0 & 372 &     545 \\
&&         &  11.5 & 382 &     558 \\

\hline
\end{tabular}
\caption{Internal conversion rates for the state $5f6d_2$ at 30223 $\mathrm{cm}^{\mathrm{-1}}$  in $~^{229\mathrm{m}}\mathrm{Th}^{+}$. The total angular momentum of this state is $J=15/2$. \label{5f6d2}}
\end{table}


\begin{table}[htb]
\begin{tabular}{|c|c||c||c|c|c|}
\hline
\multicolumn{2}{|c||}{Final state} &
\multirow{2}{*}{$E_\mathrm{m}^{\mathrm{min}}$,~eV} &
\multirow{2}{*}{$E_\mathrm{m}$,~eV} &
\multicolumn{2}{c|}{Rate, $\mathrm{s}^{-1}$} \\
\cline{1-2}
\cline{5-6}
Config. & $E$, $\mathrm{cm}^{-1}$ & &  & Pure M1 &  M1+E2 \\
\hline
\multirow{5}{*}{$7s_2$} &
\multirow{5}{*}{11961} &
\multirow{5}{*}{9.7} & 9.7 & $50813$ & 60717\\
 & & & 10.0 & $50981$                & 60913\\
 & & & 10.5 & $51252$                & 61228\\
 & & & 11.0 & $51512$                & 61532\\
 & & & 11.5 & $51761$                & 61824\\
\hline
\end{tabular}
\caption{Internal conversion rates for the state $7s_27p$ at 31626~$\mathrm{cm}^{\mathrm{-1}}$ in $~^{229\mathrm{m}}\mathrm{Th}^{+}$. The total angular momentum of this state is $J=1/2$. \label{7s27p}}
\end{table}


\subsection{Results for $\isotope{Th}^{2+}$}
With increasing charge state, the required electronic excitation is also increasing in energy. The value of $E_\mathrm{m}$ at which one can apply the same method for $\isotope{Th}^{2+}$ lies in the range from 19.1~eV to 20.0~eV (the $\isotope{Th}^{2+}$ ionization potential). Due to the relatively simple structure of the $\isotope{Th}^{2+}$ spectrum, there is no opportunity for a highly excited state to not decay through fast $E1$ transitions. However,  if $E_\mathrm{m}$ is close enough to the $\isotope{Th}^{2+}$ ionization threshold such that the excited level does not have to be very high in order to observe IC, one can expect moderate $E1$ decay rates due to their  proportionality to the third power of the photon frequency. On the other hand, we can choose a level undergoing IC at a rate high enough to compete with the $E1$ decay.

We consider two levels of $\isotope{Th}^{2+}$, namely $5f7s$ at 7501~$\mathrm{cm}^{\mathrm{-1}}$ and $6d7s$ at 16038~$\mathrm{cm}^{\mathrm{-1}}$. IC from these states becomes energetically allowed at the lowest values of $E_\mathrm{m}$ among the other states (19.1~eV and 19.2~eV, respectively).
The calculated rate of IC from the level at 7501~$\mathrm{cm}^{\mathrm{-1}}$ is equal to $7 \cdot 10^5$~$\mathrm{s}^{-1}$ and is an order of magnitude higher than the radiative decay rate $5 \cdot 10^4$~$\mathrm{s}^{-1}$ calculated with grasp2K, see Table \ref{lifetimes}. The rate of IC from the state at 16038~$\mathrm{cm}^{\mathrm{-1}}$ is calculated to be $2 \cdot 10^5$~$\mathrm{s}^{-1}$ and is at least one order of magnitude less than the radiative decay of $10^7$ $\mathrm{s}^{-1}$, see Tables \ref{tab5} and \ref{tab6}.  It is worth noting that the difference of angular momenta between the level $5f7s$ at 7501~$\mathrm{cm}^{\mathrm{-1}}$ and the ground state equals $4-3=1$ and, contrary to the case of $\isotope{Th}^+$, this allows a more straightforward excitation scheme.

\begin{table}[htb]
\begin{tabular}{|c|c||c||c|c|c|}
\hline
\multicolumn{2}{|c||}{Final state} &
\multirow{2}{*}{$E_\mathrm{m}^{\mathrm{min}}$,~eV} &
\multirow{2}{*}{$E_\mathrm{m}$,~eV} &
\multicolumn{2}{c|}{Rate, $\mathrm{s}^{-1}$} \\
\cline{1-2}
\cline{5-6}
Config. & $E$, $\mathrm{cm}^{-1}$ & &  & Pure M1 &  M1+E2 \\
\hline
\multirow{3}{*}{$5f$} &
\multirow{3}{*}{0} &
\multirow{3}{*}{19.1}
 &     19.1 & $696571$ & 696686 \\
 & & & 19.5 & $696652$ & 696767 \\
 & & & 19.9 & $696742$ & 696857 \\
\hline
\end{tabular}
\caption{Internal conversion rates for the state $5f7s$ at 7501 $\mathrm{cm}^{\mathrm{-1}}$ in $~^{229\mathrm{m}}\mathrm{Th}^{2+}$. The total angular momentum of this state is $J=3$.\label{tab5}}
\end{table}

\begin{table}[H]
\begin{tabular}{|c|c||c||c|c|c|}
\hline
\multicolumn{2}{|c||}{Final state} &
\multirow{2}{*}{$E_\mathrm{m}^{\mathrm{min}}$,~eV} &
\multirow{2}{*}{$E_\mathrm{m}$,~eV} &
\multicolumn{2}{c|}{Rate, $\mathrm{s}^{-1}$} \\
\cline{1-2}
\cline{5-6}
Config. & $E$, $\mathrm{cm}^{-1}$ & &  & Pure M1 &  M1+E2 \\
\hline
\multirow{3}{*}{$6d$} &
\multirow{3}{*}{9193} &
\multirow{3}{*}{19.2}
 &     19.2 & $194936$ & 194965 \\
 & & & 19.5 & $194984$ & 195013 \\
 & & & 19.9 & $195048$ & 195077 \\
\hline
\end{tabular}
\caption{Internal conversion rates for the state $6d7s$ at 16038 $\mathrm{cm}^{\mathrm{-1}}$ in $~^{229\mathrm{m}}\mathrm{Th}^{2+}$. The total angular momentum of this state is $J=2$.\label{tab6}}
\end{table}

\section{Possible methods to excite the $5f6d_2$ level of $\isotope{Th}^{+}$ at 30223 $\mathrm{cm}^{\mathrm{-1}}$}
The experimental population of the $5f6d_2$ state of $\isotope{Th}^{+}$ at 30223~$\mathrm{cm}^{\mathrm{-1}}$ is, at first sight, not straightforward due to the difference of the total angular momentum with respect to the ground state: $\frac{15}{2}-\frac{3}{2}=6$. In addition, if one were to excite from the ground state using six resonant photon transitions, one also needs to introduce a change in parity, i.e., one of these transitions would need to have $M1$ multipolarity. 
 In the following we therefore consider other possible ways to reach the level.

\begin{figure}[ht!]
\centering
\includegraphics[width=9cm]{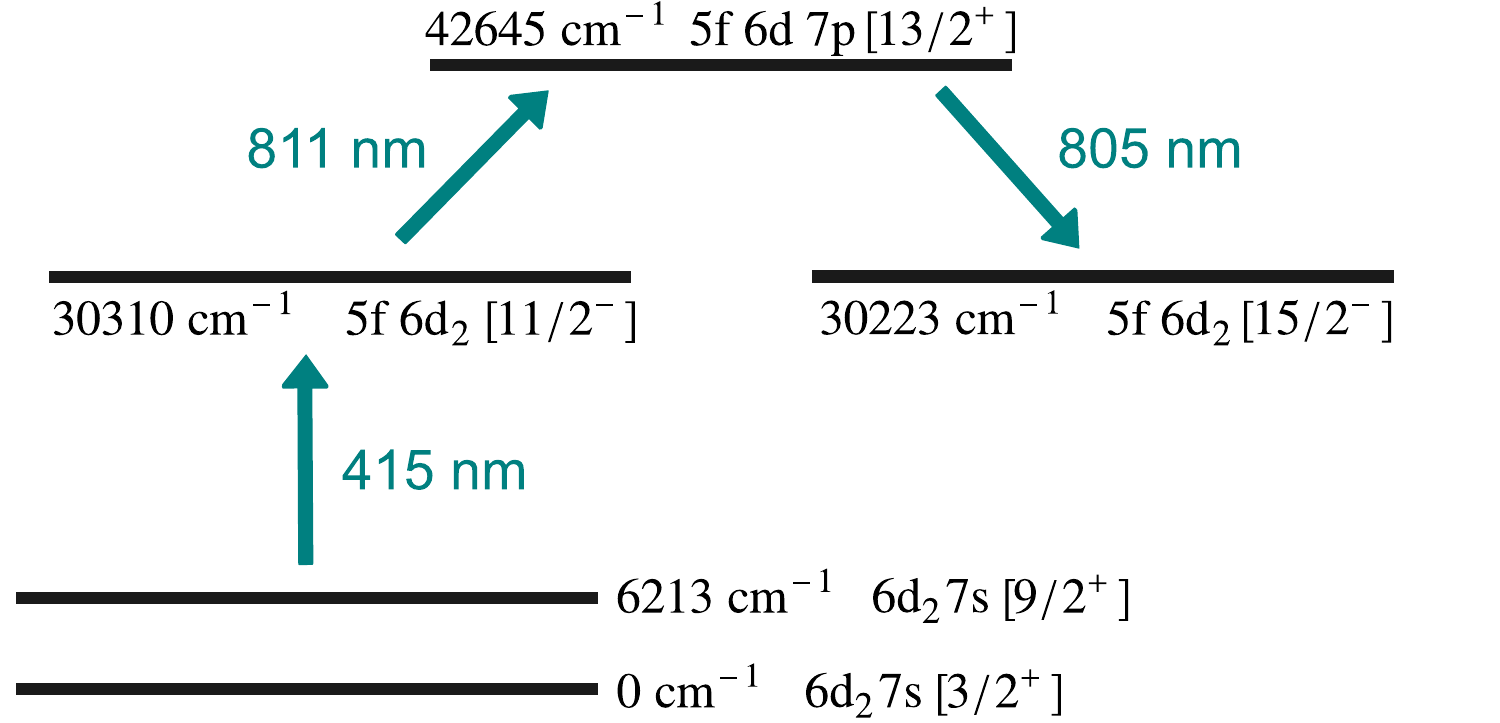}
\caption{ Possible laser excitation scheme of the $5f6d_2$ state starting from the $6d_27s$ state at 6213~$\mathrm{cm}^{\mathrm{-1}}$. \label{laserscheme}}
\end{figure}

Firstly, the state of  interest can be reached via three $E1$ excitation/de-excitation steps from the $6d_27s$ excited level at 6213~$\mathrm{cm}^{\mathrm{-1}}$ as indicated in the possible laser excitation scheme in Fig.~\ref{laserscheme}. The levels at 30310~$\mathrm{cm}^{\mathrm{-1}}$ and 42645~$\mathrm{cm}^{\mathrm{-1}}$ can serve as intermediate stages, so that one can use two laser excitations with wavelengths of  415 nm and 811 nm, followed by one induced de-excitation at 805~nm. We note that the dynamics of laser de-excitation (from Rydberg levels) has been investigated in connection with anti-hydrogen atoms \cite{Wetzels2006}. All of the wavelengths are easily produced using Ti:sapphire lasers~\cite{tisa}, one of which is frequency doubled. The $6d_27s$ state at 6213~$\mathrm{cm}^{\mathrm{-1}}$ has only one $E1$ decay channel at the extremely low photon energy of $5.6 \cdot 10^{-3}$~eV and very low decay rate of $10^{-4}$ $\mathrm{s}^{-1}$, so it can be considered as metastable.

The electronic level at 6213~$\mathrm{cm}^{\mathrm{-1}}$ has a  total angular momentum 9/2 and even parity. It can be reached with four laser (de)excitations in the following manner: an excitation from the ground state to an odd parity 3/2 level at 26965~$\mathrm{cm}^{\mathrm{-1}}$ (371~nm), a second excitation step to an even parity 5/2 state at 40644~$\mathrm{cm}^{\mathrm{-1}}$ (731~nm), a stimulated de-excitation to an odd parity 7/2 state at 18974~$\mathrm{cm}^{\mathrm{-1}}$ (461~nm) and finally a de-excitation to the even parity 9/2 level at 6213~$\mathrm{cm}^{\mathrm{-1}}$ (784~nm), all of which are suitable for Ti:sapphire lasers. Nevertheless, such an excitation scheme would require 4 different lasers and though possible, subsequent optical excitation from the metastable level towards the state of interest at 30223~$\mathrm{cm}^{\mathrm{-1}}$ is simply impractical.

Alternative means to populate the 6213~$\mathrm{cm}^{\mathrm{-1}}$ state will require experimental investigation and thus we simply highlight them here:
\begin{itemize}
\item Following the $\alpha$-decay of a source of $\isotope[233]{U}$ (which proceeds, following gamma-ray de-excitation, primarily to the ground state of $\isotope[229]{Th}$ with a 2\% branching to the low-energy isomeric state \cite{twoperc}) the $\isotope[229]{Th}$  recoils may be stopped in a buffer gas cell \cite{lars_nature,Sonnenschein2012}. Excitation of the 6213~$\mathrm{cm}^{\mathrm{-1}}$ state may occur via collisions in the gas cell;
\item Excitation of the nuclear isomeric state of the ion $\isotope[229]{Th}^+$ through an electron bridge scheme chosen such that the final electronic level is not the ground state, but the state at 6213~$\mathrm{cm}^{\mathrm{-1}}$ (this  would work for particular values of $E_{\mathrm{m}}$);
\item Thermal excitation to the metastable level via laser ablation of $\isotope[229]{Th}$ from a dried thorium nitrate solution deposited on a tungsten or tantalum substrate. This has recently been realized for the study of isotope shifts and hyperfine structures of resonance lines in $\isotope[229]{Th}^+$, with a low pressure argon buffer gas used to cool the ions to room temperature and quench the population of metastable states optically pumped by the laser excitation~\cite{Okhapkin2015}. In the current proposal however population of the metastable  state is a requirement.
\end{itemize}
%

\section{Proposed measurement for identification of the isomeric state \label{measurement}}
We propose to perform an experimental verification of the internal conversion from the aforementioned excited state of $\isotope[229]{Th}^+$ at the IGISOL facility, Jyv\"{a}skyl\"{a}, Finland. 
The study of thorium through laser spectroscopic techniques is an ongoing project at the facility, whereby an alternative to the direct detection of the isomeric decay would be the inference, through an optical measurement, of the atomic hyperfine structure. Due to different spins and magnetic moments of the ground- and isomeric state (the latter produced via the alpha decay of $\isotope[233]{U}$), such a measurement will allow the separation and identification of the two states \cite{Sonnenschein2012,Sonnenschein2012-2}. A description of the current facility and the historical developments leading towards the ion guide method of radioactive ion beam production are described in~\cite{Moore2013,Moore2014} and references therein.

\begin{figure*}
\includegraphics[width=\textwidth]{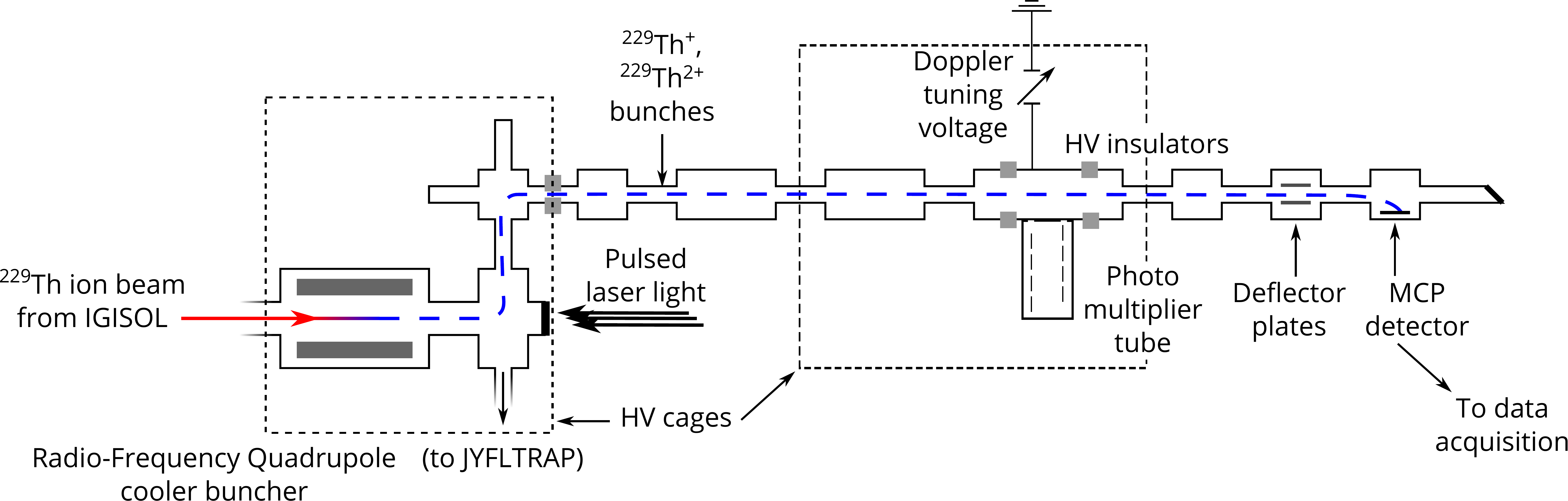}
\caption{(Color online.) Schematic diagram of the radiofrequency (rf) quadrupole cooler-buncher device connected to the collinear laser spectroscopy beamline at the \textsc{Jyfl} Accelerator Laboratory in Jyv\"{a}skyl\"{a}. Relevant details for the isomeric state detection of $\isotope[229]{Th}^{+}$ are discussed in the text.  \label{laserline}}
\label{fig:laserline}
\end{figure*}

Figure~\ref{laserline} illustrates the layout of the facility of specific relevance to our proposal. Following population of the metastable level at 6213~$\mathrm{cm}^{\mathrm{-1}}$ in $\isotope[229]{Th}$ (see previous section), the $\isotope[229]{Th}^+$ ion beam is injected into the radiofrequency (rf) quadrupole cooler-buncher~\cite{Nieminen2001}. Inside this device, the ions lose their residual energy through viscous collisions with low pressure ($\sim$1 mbar) helium gas, and a weak axial field is applied to the segmented electrodes in order to guide the ions to the exit region within 1 ms. Here, the ions may be accumulated with a trapping potential and are bunched before extraction through a miniature quadrupole into a low-energy transfer line operating at 800 V before re-acceleration by the platform potential ($\sim$30 kV) towards the experimental setups. The time structure of a typical ion bunch has been determined to be $\sim$10 $\mu$s.

In recent years, a method of optical pumping within the rf cooler-buncher in connection with collinear laser spectroscopy has been pioneered in Jyv\"{a}skyl\"{a}~\cite{Cheal2009}. Generally, laser spectroscopy is performed using electronic transitions from the ground state due to reasons of population. In order to access a wider number of elements for nuclear structure interest, excitation of ground states within the cooler-buncher using laser light generated from high power pulsed tunable Ti:sapphire lasers results in optical pumping and efficient redistribution of the ground state population to a selected metastable state. Importantly, the state survives extraction from the cooler, acceleration and delivery to the collinear set-up, the general beamline layout of which is shown in Fig.~\ref{laserline}.

In the near future, much effort will be directed towards the production of pure beams of a single isotope or even isomer. This will be realized through a multi-step excitation process with typically three pulsed lasers being used to resonantly ionize a selected element from a singly-charged to a doubly-charged state while inside the cooler, effectively an extension of the optical manipulation used to populate selected metastable states. This method is similar to that proposed for the population of the $5f6d_2$ state of $\isotope{Th}^{+}$ at 30223~$\mathrm{cm}^{\mathrm{-1}}$ from the metastable 6213~$\mathrm{cm}^{\mathrm{-1}}$ state. When the ions are released from the cooler in bunches, their time-of-flight to the detection region (a multi-channel plate, MCP, detector located at the end of the beamline) depends on the mass-to-charge ratio, $m/q$. Doubly-charged ions, in this instance $\isotope[229]{Th}^{2+}$ which is produced via IC from the excited 30223~$\mathrm{cm}^{\mathrm{-1}}$ state, will leave the cooler more quickly. The doubly-charged ion bunch will rapidly become spatially separated from the bunch of contaminant singly-charged ions thus allowing the latter to be electrostatically deflected from the beam path.

The typical flight time from the ion beam cooler-buncher to the MCP detector shown in  Fig.~\ref{laserline} for a mass $A$=100 singly-charged ion is 100 $\mu$s. This flight time scales with the square root of $m/q$ and thus for singly-charged $\isotope[229]{Th}^+$ the expected flight time increases to  approx.~1.5 ms. The corresponding flight time for doubly-charged $\isotope[229]{Th}^{2+}$ is approx.~1.1 ms. This difference in the ion bunch arrival time is sufficient for a simple electrostatic deflection of one species and therefore a clear identification of the presence of the isomeric state can be made. It is expected that background doubly-charged ions may be formed inside the cooler-buncher via resonant excitation and ionization of singly-charged thorium. This process will be studied in detail using stable $\isotope[232]{Th}^+$ which does not possess any nuclear states at optical energies.

\section{Summary and Conclusions \label{summary}}
The low-lying isomeric state of $\isotope[229]{Th}$ opens the possibility to observe for the first time IC from excited electronic states. We have investigated how this process can be used for an estimate of the nuclear transition energy and an experimental  determination of the nuclear transition strength. The scenario put forward involves laser excitation of the electronic shell in thorium ions, which subsequently opens the IC decay channel of the isomeric state. Experimentally, the decay of the nuclear excited state could be then observed via the appearance and detection of ions belonging to the next charged state. Numerical results for IC rates from excited states of $\isotope[229]{Th}^{+}$ and $\isotope[229]{Th}^{2+}$ ions have been presented. A possible experimental setup for the determination of the isomeric state properties using IC from excited electronic states at the IGISOL facility, Jyv\"{a}skyl\"{a}, Finland, has been discussed. The main challenge here appears to be the laser excitation of the electronic shell to states which live long enough to allow for IC to efficiently depopulate the isomeric state. In addition, we note that  new laser spectroscopy data for $\isotope[229]{Th}^{+}$ and $\isotope[229]{Th}^{2+}$ electronic excited states above 60000 cm$^{-1}$ might enlarge the choice of states that can be used for the proposed scheme, if additional longer-lived excited electronic states are found. Further experimental data at higher excitation energies is also important in the search for an electron bridge mechanism for the isomeric state and further laser spectroscopy scans are continuing  at the moment.


\begin{acknowledgments}
The authors gratefully acknowledge funding by the EU FET-Open project 664732. 

\end{acknowledgments}

\end{document}